\documentclass[12pt]{iopart}

\usepackage{graphicx}
\usepackage{overpic}
\usepackage{float} 
\usepackage{bbm}
\usepackage{amssymb}
\usepackage{iopams}

\newcommand{\eqref}[1]{(\ref{#1})}

\begin{document}

\title{Maxwell's demon in the quantum-Zeno regime and beyond}

\author{G. Engelhardt }

\address{Institut f\"ur Theoretische Physik, Technische Universit\"at Berlin, Hardenbergstr. 36, 10623 Berlin, Germany }
\ead{georg@itp.tu-berlin.de}

\author{G. Schaller}

\address{Institut f\"ur Theoretische Physik, Technische Universit\"at Berlin, Hardenbergstr. 36, 10623 Berlin, Germany }

\vspace{10pt}
\begin{indented}
\item[]October 2017
\end{indented}

\begin{abstract}
The long-standing paradigm of  Maxwell's demon is till nowadays a frequently investigated issue, which still provides interesting insights into basic physical questions. Considering a single-electron transistor, where we implement a Maxwell demon by a piecewise-constant feedback protocol, we investigate quantum implications of the Maxwell demon. To this end, we harness a dynamical coarse-graining method, which provides a convenient and accurate description of the system dynamics even for high measurement rates. In doing so, we are able to investigate the Maxwell demon in a quantum-Zeno regime leading to transport blockade. We argue that there is a  measurement rate providing an optimal performance. Moreover, we find that besides building up a chemical gradient, there can be also a regime where additionally the system under consideration provides energy to the demon due to the quantum measurement.
\end{abstract}

%
%
%
%
%

\section{Introduction}
 
      \begin{figure}[t]
        \centering
        \includegraphics[width=0.6\linewidth]{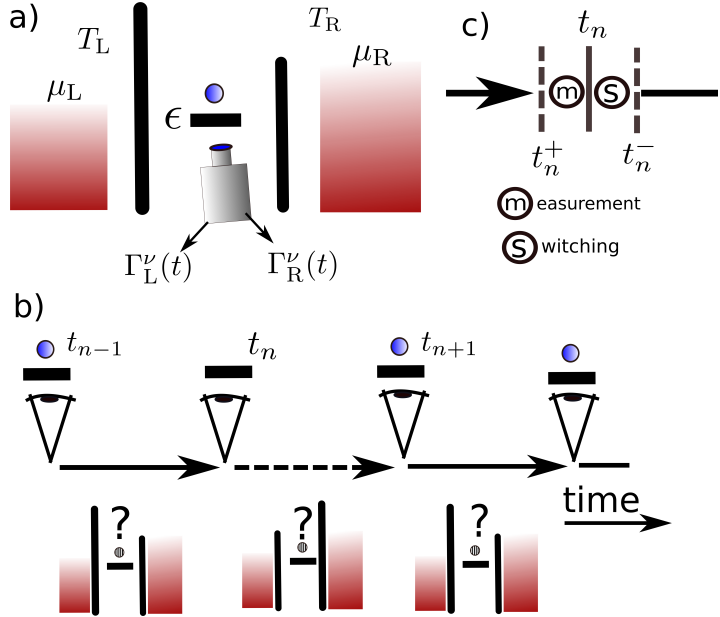}
        \caption{(a) Sketch of the SET under the action of the Maxwell demon. Two reservoirs, which are locally in  thermal equilibrium, are connected via a quantum dot with on-site energy $\epsilon$. The Maxwell demon monitors the occupation of the dot and adjusts the tunnel barriers  $\Gamma_{\alpha}^\nu(t)$ according to its observation. In doing so, one can generate a current against the chemical potential bias even at equal temperatures.
        (b) As an implementation of the Maxwell demon, we consider  a piecewise constant feedback protocol. At times $t_n= n\tau$, we projectively measure the dot occupation $\nu=\rm E,\rm F$ (empty, filled) using, e.g., a quantum-point contact. Subsequently, we adjust the tunnel rates accordingly for the next feedback period $(t_n,t_{n+1})$. On the level of equations we model this with the propagators $\exp\left[\boldsymbol{\mathcal L}^\tau_\nu \tau\right]$ as explained in Eq.~\eqref{eq:feedbackPropagator}. 
        (c) Temporal sketch of the feedback action in detail. The times $t_n^-,t_n,t_n^+$ denote the times (infinitesimally) before the measurement, after the measurement and after switching the tunnel rates, respectively.}
        \label{fig:overview}   
        \end{figure}

 Maxwell's demon is the central character in a long-standing gedankenexperiment suggested by Maxwell in 1871, which challenges the validity of the second  law of thermodynamics~\cite{Maruyama2009,Leff2014}: A box containing an ensemble of particles is divided in two compartments. The Maxwell demon observes the particles and has the ability to open and close a door such that only fast particles can enter the "left" compartment, while slow particles leave the "left" compartment. In doing so, the demon can build up a thermal gradient, which can later be used to run a thermal engine. However, as the opening and closing does not consume energy in the ideal case, this procedure would violate the second law of thermodynamics saying that such a perpetuum mobile of second kind is not possible.
 This paradox was  resolved  by Landauer by recognizing that the Maxwell demon has to delete  information  in order to perform its task. This is directly related to heat dissipation~\cite{Szilard1929, Brillouin1951,Bennett1982,Landauer1961}.

 This article provides a  quantum mechanical treatment of Maxwell's demon. In quantum mechanics the action of the Maxwell demon is more involved due to the special meaning of the observation or measurement of the particles, namely the wave-function-collapse postulate. A quantum measurement thus does not leave the system state unaffected so that the observation by the Maxwell demon necessarily affects the dynamics. Here we investigate these quantum implications  at the example of  a single-electron transistor (SET).  We implement the Maxwell demon using a  piecewise-constant feedback scheme, where the system state is observed after time periods $\tau$ and the system parameters are adjusted accordingly. This feedback scheme has been already successfully implemented in  experiment~\cite{Chida2017}. For other experimental and theoretical feedback-related approaches in mesoscopic devices to extract work or similar objective  we refer to Refs.~\cite{Strasberg2013,Strasberg2014,Kutvonen2016,Rossello2017,Strasberg2017,Schaller2011,Averin2011,Kish2012,Esposito2012,Bergli2013,Sothmann2015,Sanchez2011,Hartmann2015,Koski2014,Strasberg}.
 
 In this article, we harness a dynamical coarse-graining (DCG) method~\cite{Schaller2008}. This method provides interesting properties which are perfectly suitable for the issues which we are interesting in. 
  The DCG is designed in a way so that it becomes exact for short evolution times in contrast to a Born-Markov master equation or other coarse-graining approaches~\cite{Schaller2014a,Lidar2001,Majenz2013}. For this reason, it is favorable to use it to describe  the piecewise-constant feedback protocol for high measurement rates, as in this case the time-evolution is repeatedly restarted after each measurement. 
  
 In contrast to other methods as, e.g., the so-called Redfield equation, the DCG ensures complete positivity for all times~\cite{Schaller2008,Schaller2009,Zedler2009}, so that  thermodynamic quantities, e.g., the system entropy, are guaranteed to be well defined for all time instants. Moreover, Ref.~\cite{Rivas2017} shows that this technique even accounts for highly non-Markovian effects observable in the coherence or the entanglement dynamcis.
 Furthermore, this approach can be amended for a full-counting statistics treatment in  systems under non-equilibrium conditions~\cite{Schaller2009a}.
 
 The  DCG  thus provides a reliable accuracy for the parameter range which we are interested in. This article goes thus beyond the treatment in Ref.~\cite{Schaller2011}, where the time dynamics has been approximated by a Born-Markov master equation.

 While investigating short feedback times, we unavoidably run into another long-standing paradox of physics, namely the quantum-Zeno effect. Strictly following the principles of quantum mechanics one finds that the dynamics of a quantum system freezes when continuously measuring it with projective measurements~\cite{Misra1977,Itano1990,Fischer2001}. This paradigm is particularly interesting in the context of the classical Maxwell demon, who continuously observes the system of its interest. Indeed, we find with the system and methods at hand that the action of the Maxwell demon results in a blocking of the particle and heat currents between the reservoirs.

 Moreover, due to the action of the demon in the quantum regime, we observe another side effect. Besides building up a chemical potential gradient between the two reservoirs which could be used to charge a battery, we argue that there can be also a net energy decrease of the system due to the feedback action. We explain that it is most convenient to run the Maxwell demon in such a regime, as we do not have to invest external power in order to make the demon work.
 
 This article is organized as follows: In Sec.~\ref{sec:modelAndMethods}, we explain the SET, which we describe by a Fano-Anderson model. We give a compact introduction to the the DCG method applied throughout the article and prove its validity. In Sec.~\ref{sec:feedbackControl}, we explain the implementation of the Maxwell demon by a piecewise-constant feedback scheme and show how to model this on the level of the equation of motion for the reduced density matrix. We show how the DCG approach reveals the quantum-Zeno effect for a continuous measurement. In Sec.~\ref{sec:powerGainHeatFlow}, we discuss the thermodynamic properties of the system like electric power, gain, heat flow and entropy production in the quantum-Zeno regime and beyond. In Sec.~\ref{sec:discussionsConclusions}, we provide a concluding discussion of our results. Supplemental information is given in the appendix.

 \section{Model and  methods}
 \label{sec:modelAndMethods}

 We implement the Maxwell demon in a SET, which is a mesoscopic transport setup consisting of two electron reservoirs coupled by a quantum dot. A sketch of the system is depicted in Fig.~\ref{fig:overview}(a).
   We model the SET by a two-terminal Fano-Anderson model, whose Hamiltonian reads~\cite{Gurvitz1991,Topp2015,Engelhardt2016}
 \begin{eqnarray}
 	\hat H^\nu =&\hat H_{\rm d} + \hat H_{\rm r} + \hat H_{\rm c}^\nu \label{eq:totalHamiltonian},\\
 	\hat H_{\rm d}=& \epsilon \hat c_{\rm d}^\dagger \hat c_{\rm d},\nonumber\\
 	\hat H_{\rm r}=&\sum_{k,\alpha=\rm R, L} \omega_{k,\alpha} \hat c_{k,\alpha}^\dagger \hat c_{k,\alpha} \nonumber = \sum_{\alpha=\rm R, L}H_{\rm r,\alpha}, \\
 	\hat H_{\rm c}^\nu=& \sum_{k,\alpha=\rm R,\rm L } t_{k,\alpha}^\nu  \left( \hat c_{\rm d}^\dagger \hat  c_{k,\alpha} + \mathrm h.c.  \right)= \sum_{\alpha=\rm R, L}H_{\rm c,\alpha}^\nu,\nonumber
 \end{eqnarray}
 where $\hat c_{\rm d}^\dagger$ and $\hat c_{k,\alpha}^\dagger$ are  fermionic operators representing the central dot with on-site energy $\epsilon$ and the reservoir states with energies $\omega_{k,\alpha}$, respectively. Thereby, $\alpha=\rm R, L $ (right, left) labels the reservoirs  and $k$ are their internal states. The hopping amplitudes between dot and reservoir states are given by $ t_{k,\alpha}^\nu$. 
      
 We have introduced the index $\nu$ to implement a feedback protocol: the Hamiltonian (or more precisely the hopping amplitudes) will be conditioned on the dot occupation $\nu=\rm E,F$ (empty, filled). We provide more details in Sec.~\ref{sec:feedbackControl}.
 
  The initial condition which we consider throughout the article is given by
 \begin{eqnarray}
 \rho(0)&=\rho_{\rm d}^0\otimes \rho_{\rm L}(0) \otimes \rho_{\rm R}(0) \label{eq:initialState} , \\
 &\rho_{\alpha}(0)= \frac{1}{Z_\alpha}e^{-\beta_\alpha \left(\hat H_{\rm r,\alpha}-\mu_{\alpha} \hat N_{\rm r,\alpha} \right)},
\nonumber 
 \end{eqnarray}
 where $\rho_{\rm d}^0 $ is the initial density matrix of the dot and $\rho_{\rm \alpha}(0)$ are the initial density matrices of the reservoirs.
 Thus, the reservoirs are considered to be locally in a thermal equilibrium state with inverse temperatures $\beta_\alpha= 1 /(k_{\rm B} T_\alpha)$ and chemical potentials $\mu_\alpha$. Here, $Z_\alpha$ denotes the partition function which ensures that $\rm{Tr}\left[\rho_{\alpha}(0)\right]=1$ and $\hat N_{\alpha} = \sum_k \hat c_{k\alpha}^\dagger \hat c_{k\alpha} $ is the particle number operator of the reservoir $\alpha$.

 In the following, we apply a DCG method~\cite{Schaller2009} to calculate the dynamics~ of the reduced density matrix of the quantum dot $\rho_{\rm d}(t)= \rm{Tr}_{\rm r}\left[ \rho(t)\right]$, where $\rm{Tr}_{\rm r}\left[ .\right]$ denotes the trace over the reservoir degrees of freedom. We represent the reduced density matrix of the quantum dot in the local basis $\left|0\right>=\left|\rm {vac}\right>$ and $\left|1\right>=\hat c_{\rm d}^\dagger\left|\rm {vac}\right>$ and introduce the notation $\rho_{\rm d,\lambda \lambda'}\equiv \left<\lambda \right| \rho \left|\lambda'\right>$. In doing so, the diagonal elements $\underline \sigma=(\rho_{\rm d,00},\rho_{\rm d,11})$ of the reduced density matrix of the system decouple from the coherences and approximately read as a function of time
 \begin{equation}
 \underline \sigma( t) = \exp\left[ {\sum_{\alpha=\rm R,\rm L}\boldsymbol{\mathcal L}^{ t}_{\alpha,\nu}\left(\underline \xi \right)  t} \right]\underline \sigma(0),
 \label{eq:TimeEvolutionCG}
 \end{equation}
 where the coarse-grained Liovillian reads
 \begin{equation}
 	\mathbf L^t_{\alpha,\nu}\left(\underline \xi \right)=
 	 \left(
 	\begin{array}{cc}
 		 -\gamma_{10}^{t,\alpha,\nu}(0)  & \gamma_{01}^{t,\alpha,\nu} (\zeta_\alpha) e^{ \rm i \chi_\alpha } \\
 		 \gamma_{10}^{t,\alpha,\nu}(\zeta_\alpha)e^{- \rm i \chi_\alpha }  & - \gamma_{01}^{t,\alpha,\nu}(0)  \\
 		 \end{array}
 	\right).
 	\label{eq.cgLiovillian}
 \end{equation}
 In Eq.~\eqref{eq.cgLiovillian} we additionally introduced the counting fields $\chi_\alpha$ and $\zeta_\alpha$, which allow for a determination of the number of  particles $\Delta n_{\alpha}$ and the amount of energy  $\Delta E_{\alpha}$ entering the reservoir $\alpha$ during the time interval $t'\in \left(0,t\right)$. For brevity we thereby combine the counting fields in the (transposed) vector  $\underline \xi ^T =(\chi_{\rm L},\chi_{\rm R},\zeta_{\rm L},\zeta_{\rm R})$. The matrix entries read 
\begin{eqnarray}
	\gamma_{10}^{t,\alpha,\nu}(\zeta_\alpha)  &= \frac{t}{2 \pi} \int d\omega  {\rm sinc} ^2 \left[ \frac t 2 \left( -\epsilon -\omega \right)  \right] \gamma_{10}^{\alpha,\nu}(\omega)  e^{i \zeta_\alpha \omega}, \\
	\gamma_{01}^{t,\alpha,\nu}(\zeta_\alpha) &= \frac{t}{2 \pi}  \int d\omega  {\rm sinc}^2 \left[ \frac t 2 \left( \epsilon -\omega \right)  \right] \gamma_{01}^{\alpha,\nu}(\omega) e^{i \zeta_\alpha \omega} ,\nonumber
\end{eqnarray}
where ${\rm sinc}(x)\equiv \sin(x)/x$ is the sinc function and we have used the abbreviations 
 \begin{eqnarray}
 	\gamma_{10}^{\alpha,\nu}(\omega) &=  \Gamma_{\alpha}^{\nu}(-\omega) f_\alpha(-\omega), \\
 	\gamma_{01}^{\alpha,\nu}(\omega) &=   \Gamma_{\alpha}^{\nu}(\omega)\left[ 1- f_\alpha(\omega) \right] ,
 \end{eqnarray}
 with the Fermi function \mbox{$f_\alpha(\omega) = 1 /(e^{\beta_\alpha(\omega-\mu_\alpha)}+1)$} and the spectral coupling density \mbox{$\Gamma_\alpha(\omega) = \sum_{k}\left| t_{k,\alpha}^\nu\right|^2\delta(\omega-\omega_{k,\alpha})$}.
 
 In the numerical calculations throughout the article, we use
 \begin{equation}
 \Gamma_{\alpha}^{\nu} (\omega)= \Gamma_{0,\alpha}^{\nu} \frac{\delta_\alpha^2 \Theta(\omega-\omega_{\rm min})\Theta(\omega_{\rm max}-\omega) }{(\omega-\epsilon_\alpha)^2+\delta_\alpha^2 } .
 \label{eq:spectralCouplingDensity}
 \end{equation}
 This is a Lorentz function which is centered around $\epsilon_\alpha$ and has a width $\delta_\alpha$. The function $\Theta(x)$ is the Heavyside function which ensures a compact support of the spectral coupling density between $\omega_{\rm min}$ and $\omega_{\rm max}$ needed for numerical calculations.

\begin{figure}[t]
 	\centering
 	\includegraphics[width=0.65\linewidth]{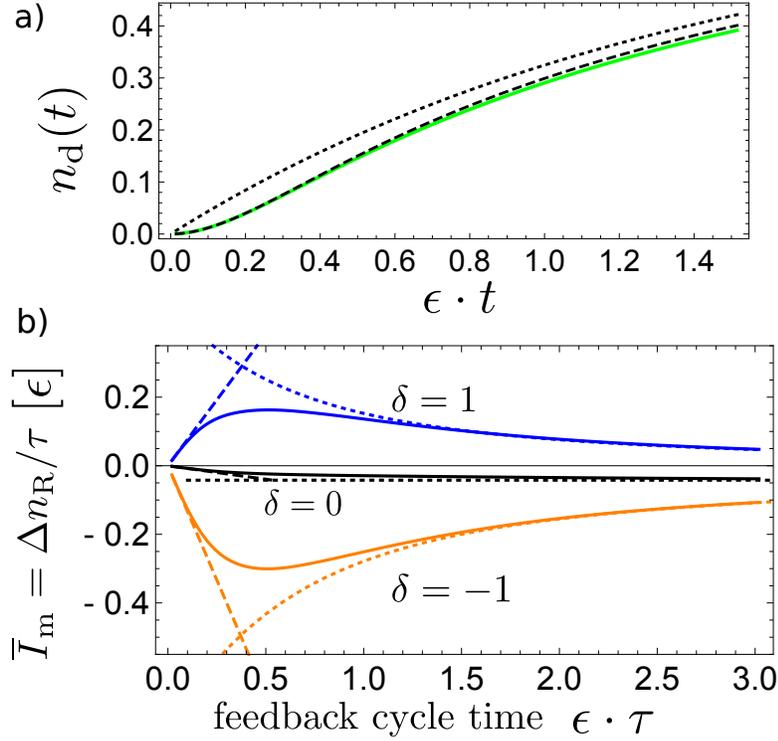}
 	\caption{(a) Dot occupation $n_{\rm d}(t)$ as a function of time. The results of the DCG approach, the exact solution and the BMS master equation are depicted with a solid (green), dashed and dotted line, respectively. The parameters are $\Gamma_{0, \alpha }^{\nu }=0.5$, $\epsilon_{\rm L}=5\epsilon $, $\epsilon_{\rm R}=-\epsilon$, $\delta_\alpha=5 \epsilon$, $\omega_{\rm{max}}=+\infty$, $\omega_{\rm{min}}=-\infty $, $\mu_{\rm L} =0$,  $\mu_{\rm L} =10 \epsilon$  and $T_{\rm L}=T_{\rm R}= 10\epsilon$.
 		(b) Time-averaged currents $\overline I_{\rm m}$ under the action of the Maxwell demon in the stationary state as a function of the feedback time $\tau$ investigated in Sec.~\ref{sec:feedbackControl}. Overall parameters are as in panel (a) with $\Gamma_{0,\alpha}^\nu=0.5\epsilon $, except that we have changed the cut-off frequencies to $\omega_{\rm{max} }=20 \epsilon$, $\omega_{\rm{min} }=0$. The curves with feedback parameter $\delta=-1,0,1$ are depicted in blue, orange and black, respectively. The solid, dashed and dotted lines depict the solution with the DCG method, the linear expansion for short times $\tau$ and the BMS master equation result, respectively.
  }
 	\label{fig:fbCurrent}   
 \end{figure}

 In Fig.~\ref{fig:fbCurrent}(a), the accuracy of the DCG method is benchmarked  against the exact solution of the Fano-Anderson model in the absence of feedback action~\cite{Topp2015}. There, we depict the occupation of the dot, which is given by $n_{\rm d}(t)= \rho_{\rm d,11}(t)$. The DCG approach is optimized to resemble the exact dynamics for short times $t$~\cite{Schaller2009,Schaller2008}, see Fig.~\ref{fig:fbCurrent}(a). By construction, in the long-time limit $t\rightarrow\infty$ the DCG dynamics converges to the dynamics of the Born-Markov-Secular (BMS) master equation, which resembles the exact solution for the parameters under consideration. Consequently, the DCG method guarantees a good performance for short times in all parameter regimes and for long times in the weak-coupling limit.
 Importantly, due to its construction, the DCG method maintains a Lindblad form in Eq.~\eqref{eq.cgLiovillian} for all times $t$, which ensures positivity of $\rho_{\rm d}$ and consequently guarantees well-defined thermodynamic calculations.

 \section{ Feedback control}
 \label{sec:feedbackControl}
 
  In order to implement the Maxwell demon we apply a projective measurement in combination with a piecewise-constant feedback scheme. This is sketched in Fig.~\ref{fig:overview}(b). At times $t_n=n \tau$ we conduct projective measurements of the dot occupation. According to the outcome, we adjust the system parameters which then remain constant for the following time interval $t\in (t_n,t_{n+1})$. In particular, here we vary the tunnel barriers which are parameterized by $\Gamma_\alpha^\nu(\omega)$ with $\nu=\rm E,\rm F$ if the dot occupation has been empty or filled  at time $t_n$, respectively.  
  
  \subsection{Action on the density matrix of the total system}
  
  \label{sec:fbActionTotalDensityMatrix}
  The feedback interventions occur at times $t_n=n \tau$. Within the time intervals $\left(t_n , t_{n+1} \right)$ the total system (including the reservoirs) evolves under the Hamiltonian Eq.~\eqref{eq:totalHamiltonian} conditioned on $\nu=\rm E,\rm F$ and is therefore conservative. The total energy can thus only change during the feedback interventions at times $t_n=n\tau$. 
  
  The intervention can be divided in two steps. To this end, we introduce the (virtual) times $t_n^+$ and $t_n^-$ as depicted in Fig.~\ref{fig:overview}(c) for illustration. First, one has to  measure the dot occupation. This measurement shall take place during the time interval $(t_n^-,t_n)$. Second, according to the measurement outcome, we  adjust the tunnel barriers $\Gamma_{\alpha}^\nu (\omega)$. This step shall take place in the time interval $(t_n,t_n^+)$. However, as we explain in Sec.~\ref{sec:gain}, for projective measurements the switching work can be here neglected, such that only the first step changes the  energy of the total system.
  
  As the total Hamiltonian Eq.~\eqref{eq:totalHamiltonian} is quadratic, the most important observables as the total system energy can be determined by the single-particle density matrix
  \begin{equation}
  \rho_{x,y} = \rm{Tr}\left[\hat c_{x}^\dagger \hat c_y \rho\right],
  \end{equation}
  with $x\in \left\lbrace \rm{d},(k,\alpha) \right \rbrace$. 
 
  The projectors which project the system state to the empty and filled quantum dot are given by
  \begin{equation}
  \hat P_{\rm E} = \hat c_{\rm d}\hat c_{\rm d}^\dagger, \qquad\hat P_{\rm F} = \hat c_{\rm d}^\dagger \hat c_{\rm d}.
  \end{equation}
  Their action on the total system density matrix is accordingly~\cite{Wiseman2009}
  \begin{equation}
  \rho(t_n^-)\rightarrow \rho^\nu (t_n)= \frac{\hat P_\nu \rho(t_n^-)\hat P_\nu }{\rm{Tr}\left[\hat  P_\nu \rho(t_n^-)\right]}.
  \label{eq:projectionDensityMatrix}
  \end{equation}
  In consequence, it is easy to see that the single-particle density matrix elements become $\rho_{\rm d,d}\rightarrow \left\lbrace0,1\right\rbrace$ and $\rho_{\rm d,(k,\alpha)}\rightarrow 0$.  The two distinct measurement results $\nu=\rm E,\rm F$ can be found with probabilities $p_\nu=\rm{Tr}\left[ \hat P_\nu \rho(t_n^-)\right] $, so that the density matrix of the total system after the measurement irrespective of the measurement outcome reads
  \begin{equation}
  \rho(t_n) = \sum_{\nu=\rm E,\rm F} p_n \rho^\nu(t_n) = \sum_{\nu=\rm E,F}\hat  P_\nu \rho(t_n^-)\hat P_\nu.
  \end{equation}
  In the subsequent time interval $(t_n^+,t_{n+1}^-)$, the dynamics is determined by the Hamiltonian  $\hat H^\nu$ depending on the measurement result $\nu=\rm E,\rm F$. 
  
  \subsection{Time evolution of the reduced density matrix}
  
    \begin{figure*}[t]
      \centering
      \includegraphics[width=1\linewidth]{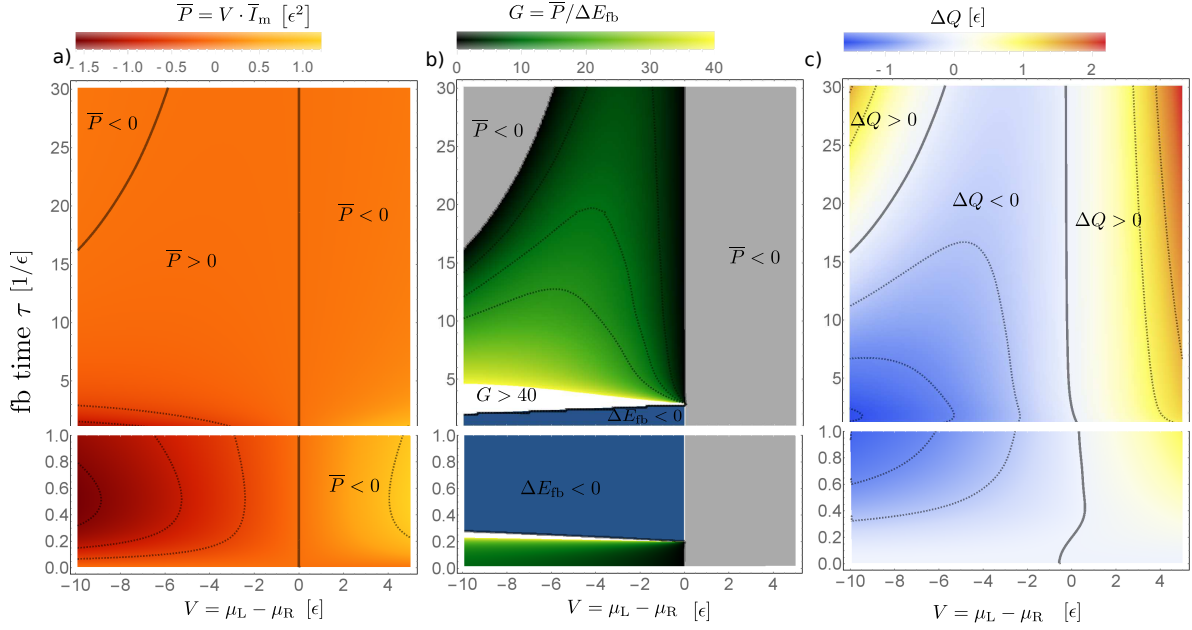}
      \caption{(a) Time-averaged power $\overline P$ as a function of the chemical potential bias $V$ and the feedback time $\tau$. Dashed lines depict sets of equal power. The solid lines mark the set with $\overline P=0$. Overall parameters are as in Fig.~\ref{fig:fbCurrent}(b) with $\delta=1$.
      (b) depicts the corresponding gain defined in Eq.~\eqref{eq:defGain}. In the gray region we find $\overline P<0$ and the gain is not interesting as we waste power. In the white region we find $G>40$. The  most efficient way to run the Maxwell demon is in the blue region, where $\Delta E_{\rm fb}<0$ so that one does not have to  invest power to run the feedback protocol.
  (c) Total amount of heat $\Delta Q$ defined in Eq.~\eqref{eq:defHeat} which enters ($\Delta Q>0$) or leaves ($\Delta Q<0$) the reservoirs. The solid lines mark the set with $\Delta Q=0$. The regions of $\Delta Q<0$ are strongly correlated to the regions of $\overline P>0$. }
      \label{fig:powerDiagram}   
      \end{figure*}
  
  In the following, we describe the dynamics on the level of the reduced density matrix of the quantum dot $\rho_{\rm d}$. For its diagonal elements contained in the vector $\underline \sigma$, the projective  measurement in Eq.~\eqref{eq:projectionDensityMatrix} translates into
  \begin{equation}
  \hat P_\nu \rho \hat P_\nu \quad \hat{=}\quad \boldsymbol{\mathcal P}_\nu \underline \sigma,
  \end{equation}
 where
 \begin{equation}
 \boldsymbol {\mathcal P}_{\rm E} = \left(
 	\begin{array}{cc}
 		 1 &  0\\
 		 0  &  0 \\
 		 \end{array}
 	\right),
 	\qquad \qquad
 \boldsymbol {\mathcal P}_{\rm F} = \left(
 	\begin{array}{cc}
 		 0 &  0\\
 		 0  &  1 \\
 		 \end{array}
 	\right).
 \end{equation}
are the  measurement operators corresponding to the measurement results $\nu=\rm E,F$.

The projective measurement is subsequently followed by a time evolution which is conditioned on the measurement result. The conditioned time evolution  can be described by the DCG approach  in Eq.~\eqref{eq:TimeEvolutionCG},  so that  the feedback time-evolution propagator  reads~\cite{Schaller2012}
 \begin{equation}
 	\boldsymbol{\mathcal F}^\tau (\underline \xi) = e^{\boldsymbol {\mathcal L}^\tau_{\rm E}(\underline\xi)\tau} \boldsymbol {\mathcal P}_{\rm E}  + e^{\boldsymbol {\mathcal L}^\tau_{\rm F}(\underline \xi)\tau} \boldsymbol {\mathcal P}_{\rm F},
 	\label{eq:feedbackPropagator}
 \end{equation}
 where $\boldsymbol {\mathcal L} _\nu^\tau (\underline \xi) =\sum_{\alpha} \boldsymbol {\mathcal L} _{\alpha,\nu}^\tau(\underline \xi)$.
 The propagator $\boldsymbol {\mathcal F}^\tau (\underline \xi)$ evolves the reduced density matrix by one feedback period $\tau$.  
 
 From the generalized propagator $\boldsymbol {\mathcal F}^\tau (\underline \xi)$ we obtain the moment generation function (MGF)
 \begin{equation}
 M(\tau,\underline \xi)  = \left( 1,1 \right)\boldsymbol {\mathcal F}^\tau (\underline \xi) \underline\sigma_{\rm s},
 \end{equation}
where $\underline \sigma_{\rm s}$ denotes the stationary density matrix, as we are interested in the long-term dynamics. The stroboscopic stationary state at times $t_n=n \tau$ is the eigenvector of $\boldsymbol {\mathcal F}^\tau(0)$ with eigenvalue $ \varphi_1=1$,
 \begin{equation}
 \boldsymbol {\mathcal F}^\tau(0)\underline \sigma_{\rm s} =\underline \sigma_{\rm s},
 \end{equation}
 which always exists and depends on the measurement rate $\tau$. Furthermore, it can be shown that the second eigenvalue fulfills $0\leq\varphi_2\leq 1$. It thus describes the relaxation dynamics towards the stationary state. In terms of the MGF, the number of particles entering reservoir $\alpha$ within the time interval $\tau$  is given by~\cite{Schaller2014a}
 ~
 \begin{eqnarray}
 \Delta n_\alpha &=- \rm i \frac{d}{d \chi_\alpha} \left. M(\tau,\underline \xi) \right|_{\underline \xi=0} .
 \end{eqnarray}
 In the same way we obtain the change of energy $\Delta E_\alpha$ in reservoir $\alpha$ by deriving $ M(\tau,\underline \xi)$ with respect to $\zeta_\alpha$ instead.

 In Fig.~\ref{fig:fbCurrent}(b), we depict the time-averaged matter current $\overline I_{\rm m}= \Delta n_{\rm R} /\tau= -\Delta n_{\rm L} /\tau $ as a function of $\tau$. In doing so, we have chosen the parametrization of the spectral coupling density as in Eq.~\eqref{eq:spectralCouplingDensity}, but with the proportional parameter adjusted to
 \begin{eqnarray}
 \Gamma_{0,\rm L}^\nu&\rightarrow 
 \left\lbrace
 \begin{array}{cc}
 \Gamma_{0,\rm L} e^{+\delta} &\nu =\rm E\\
 \Gamma_{0,\rm L} e^{-\delta} &\nu =\rm F,
 \end{array}
 \right.\nonumber \\
 \Gamma_{0,\rm R}^\nu&\rightarrow 
 \left\lbrace
 \begin{array}{cc}
 \Gamma_{0,\rm R} e^{-\delta} &\nu =\rm E\\
 \Gamma_{0,\rm R} e^{+\delta} &\nu =\rm F
 \end{array}
 \right. .
 \end{eqnarray}
 The parameter $\delta$ controls the feedback action. 
 For $\delta=0$, there is no feedback control. For $\delta>0$, the feedback control supports a matter current from the left to the right reservoir, while for $\delta <0$ the feedback supports the opposite direction.
 
We depict $\overline I_{\rm m}$ for three different feedback strengths $\delta$ with solid lines. Here and in the following, we choose equal temperatures $T_{\rm L}=T_{\rm R}=T$ in order to exclude a thermoelectric effect~\cite{Topp2015}. For the chemical potentials we consider $\mu_{\rm L}<\mu_{\rm R}$. Consequently, the time-averaged current $\overline I_{\rm m}$ is negative in the absence of feedback, as can be found for $\delta=0$. For a positive feedback parameter $\delta=1$, we can find a current against the bias $V=\mu_{\rm L}-\mu_{\rm R}$ and we generate a time-averaged electric power, which we define as
 \begin{equation}
\overline  P \cdot \tau= - \Delta n_{\rm R} \cdot V.
\label{eq:timeAveragedBias}
 \end{equation}
 Using this definition, we generate electric power for $\overline P>0$ and waste power for  $\overline P<0$. For $\delta<0$, our numerical calculations verify that the feedback protocol  supports the  current along the chemical potential bias. 
 
 For long feedback times $\tau\rightarrow \infty$ the time-averaged current is always directed along the chemical potential bias irrespective of the feedback strength $\delta$. Consequently, for positive feedback strength $\delta$ which implies a current against the bias for rather short $\tau$, there must be a $\tau_0$ at which the time-averaged current vanishes, thus $\overline I_{\rm m}(\tau_0)=0$. This can be explained as follows. In the limit of long feedback times $\tau$, the dynamics of the propagators conditioned on $\nu=\rm{E,F}$  in Eq.~\eqref{eq:feedbackPropagator} converges to the ones of the BMS master equation, respectively~\cite{Schaller2008}. Regardless of the feedback time $\tau$, the propagator in Eq.~\eqref{eq:feedbackPropagator} describes  an average of two distinct time evolutions with no feedback. For the BMS master equation (in the absence of feedback and at equal temperatures) it is known that  the current  always flows along the chemical potential bias in the long-time limit. This is a consequence of the second law of thermodynamics which is respected by the BMS master equation. Consequently, the measurement-averaged current becomes directed along the chemical potential gradient.

 \subsection{Maxwell demon in the quantum-Zeno regime}
 
In  the limit of continuous feedback $\tau\rightarrow 0$, the current vanishes independent of the feedback parameter $\delta$ as can be observed in Fig.~\ref{fig:fbCurrent}(b). This can be explained with the quantum-Zeno effect. For $\tau=0$, the propagator calculated using the DCG approach in Eq.~\eqref{eq:feedbackPropagator} becomes
 \begin{equation}
 \mathcal F^0\left(\underline \xi\right) = 
 \left(
  	\begin{array}{cc}
  		 1 &  0\\
  		 0  &  1 \\
  		 \end{array}
  		 \right).
 \end{equation}
This means that now both eigenvalues are $\varphi_\lambda=1$. As a consequence, the system is now bistable: Either the dot is occupied or empty for all times. In addition, coherences are continuously projected to zero as discussed in Sec.~\ref{sec:fbActionTotalDensityMatrix}. Due to the infinite measurement rate, the dot dynamics gets thus frozen, so that no particle can enter or leave  the  reservoirs. Thus, the DCG method under consideration resembles the quantum-Zeno effect~\cite{Misra1977,Itano1990,Fischer2001}.

In order to find a compact approximation and an intuitive explanation for the behavior in the quantum-Zeno regime, we expand the MGF for short times up to the lowest non-vanishing order in $\tau$ which still contains a dependence on the counting field. In doing so, we find
\begin{eqnarray}
M(\chi,\tau) =&(1,1)\left( 1+ \boldsymbol{ \mathcal L}_{\rm E}(\chi)\tau^2  \boldsymbol{ \mathcal P}_{\rm E}  + \boldsymbol{ \mathcal L}_{\rm F}(\chi)\tau^2 \boldsymbol{ \mathcal P}_{\rm F}+\dots\right)\nonumber \\
&\left(\underline \sigma_{s,0}+\dots\right),
\label{eq:cgfZenoRegime}
\end{eqnarray}
where we have defined
\begin{equation}
	\boldsymbol{ \mathcal L}_{\nu} \left(\underline \xi\right)=
	\sum_{\alpha=\rm {R,L} }
\left(
 	\begin{array}{cc}
 		 -g_{10}^{\alpha,\nu}(0)  & g_{01,01}^{\alpha,\nu} (\zeta_\alpha) e^{ i \chi_\alpha } \\
 		 g_{10}^{\alpha,\nu}(\zeta_\alpha)e^{ -i \chi_\alpha }  & - g_{01}^{\alpha,\nu}  \\
 		 \end{array}
 	\right),
 	\label{eq:liouvillianShortTimes}
\end{equation}
and
\begin{eqnarray}
	g_{10}^{\alpha,\nu} (\zeta_\alpha) &=  \frac{1}{2  \pi} \int d\omega  \Gamma_\alpha^\nu(\omega) e^{-i\zeta_\alpha \omega} f_\alpha(\omega),\\
	g_{01}^{\alpha,\nu} (\zeta_\alpha)  &=  \frac{1}{2  \pi} \int d\omega  \Gamma_\alpha^\nu(\omega) e^{i\zeta_\alpha\omega}\left[ 1-f_\alpha(\omega)\right].
	\label{eq:coefficientsLiouvillianShortTimes}
\end{eqnarray}
We note that the spectral coupling density $\Gamma_\alpha^\nu(\omega)$ must ensure an  appropriate frequency cutoff in order to avoid that higher derivatives with respect to $\zeta_\alpha$ diverge.
The MGF in the short feedback time limit thus reads
\begin{eqnarray}
m\left(\underline \xi,\tau\right) =&\tau^2 \sum_{\alpha=\rm {R,L} } n_{\rm s,0}  g_{01}^{\alpha,\rm F} (\xi_\alpha)  e^{-i\chi_\alpha } \nonumber \\
&\qquad +  \left(1-n_{\rm s,0}  \right) g_{10}^{\alpha,\rm E} (\xi_\alpha) e^{i\chi_ \alpha} ,
\end{eqnarray}
 where
 \begin{equation}
 	n_{\rm s} 
 	\rightarrow n_{\rm s,0}= \frac{1} {1+ \frac{g_{01}^{\rm L,\rm F}(0)+g_{01}^{\rm R,\rm F} (0)}  {g_{10}^{\rm L,\rm F}(0)+g_{10}^{\rm R,\rm F}(0) }}
 \end{equation}
 denotes the corresponding occupation of the dot. We emphasize that the first non-vanishing order of the MGF is $\propto\tau^2$. In consequence,  the time-averaged current $\overline I_{\rm m}=\Delta n_{\rm R}/\tau$ and all higher cumulants vanish for $\tau\rightarrow 0$. This $\tau^2$ behavior is a typical feature in the Zeno-regime~\cite{  Itano1990}. 
 
 Generally, the BMS master equation results are not valid in the short-time regime.
 A short-time expansion as before reveals why the BMS treatment provides an inaccurate result as can be seen in Fig.~\ref{fig:fbCurrent}(b). Formally, the time evolution of the BMS approach reads as in Eq.~\eqref{eq:TimeEvolutionCG}, but with the time-dependent matrices replaced by  time-independent ones, thus $\boldsymbol{ \mathcal L}^\tau_{\alpha}\rightarrow \boldsymbol{ \mathcal L}^{\rm {BMS}}_{\alpha}$. Performing the same expansion of the MGF, we find that $M\left(\underline{\zeta},\tau\right)\propto \tau$. Consequently, the Born-Markov  treatment  leads to a finite time-averaged current $\overline I_{\rm m}$ even for vanishing feedback times.

 \section{Power, gain and heat flow}
 \label{sec:powerGainHeatFlow}
 
 \subsection{Power}
 
 In Fig.~\ref{fig:powerDiagram}(a), we depict the power as a function of the bias $V=\mu_{\rm L}-\mu_{\rm R }$ and the feedback time $\tau$. 
  The dashed lines depict levels of equal power. The solid lines show the set of $\overline P=0$. There are two ways to cross this boundary. At the line $V=0$ the bias changes sign, while in the upper left region of the diagram there is a sign change of $\overline P$ as the time-averaged current $\overline I_{\rm m}$ changes its direction. Overall, the power is close to zero in wide parts of the diagram, but shows more structure for small $\tau$.  Here, we see that the feedback scheme generates most power for large negative bias and intermediate feedback times $\epsilon \tau\approx 0.5 $. On the other hand, most power is wasted for a large positive power and an intermediate feedback time.  
  
  \subsection{Gain}
  
  \label{sec:gain}
  In order to estimate the performance of  Maxwell's demon, power is not the only decisive quantity. As the total process is not conservative, the  energy of the total system changes. In particular, we find that the total energy can increase or even decrease on average. This amount of energy change is denoted with feedback energy $\Delta E_{\rm fb}$ in the following. For $\Delta E_{\rm fb}>0$, the action of the Maxwell demon  leads to an increase of the total system energy, while for $\Delta E_{\rm fb}<0$, the total system energy decreases. We define the corresponding gain parameter 
  \begin{equation}
  G = \frac{\overline P\cdot \tau}{\Delta E_{\rm {fb}} }\cdot \Theta\left( \overline P\right)\cdot \Theta\left( \Delta E_{\rm {fb}} \right).
  \label{eq:defGain}
  \end{equation}
where we restrict the definition of the gain to positive power  $\overline P>0$ and feedback energy $\Delta E_{\rm {fb}} >0$. 
 The feedback energy can be calculated by considering in detail the feedback process  as sketched in Fig.~\ref{fig:overview}(b),(c) and explained in Sec.~\ref{sec:fbActionTotalDensityMatrix}. 
  
 We first determine the change of the mean energy of the total system due to the measurement in the virtual time interval $t\in (t_{n}^-,t_n)$. To this end, we have to determine the difference of the energies of the total system Eq.~\eqref{eq:totalHamiltonian} before and after the measurement,
 \begin{equation}
 \Delta E_{\rm{tot}}^\nu \left[ \rho(t_n^-)\right]= \rm{Tr}\left[ \hat H^\nu\rho(t_n)\right]-\rm{Tr}\left[ \hat H^\nu \rho(t_n^-)\right].
 \end{equation}
 Thus, we compare the energy of the state shortly before the measurement $\rho(t_n^-)$ with the state after the measurement $\rho(t_n)$ with regard to the total Hamiltonian before the measurement.  
 As theses density matrices  differ only in the dot-reservoir coherences $\rho_{\rm d,(k,\alpha)}$, which vanish due to the measurement $\rho_{\rm d,(k,\alpha)}(t_n)\rightarrow 0$, we find
 \begin{equation}
 \Delta E_{\rm{tot}}^\nu \left[ \rho(t_n^-)\right]= \left< \hat H_{\rm c}^\nu\right>_{t_n^-},
 \end{equation}
 where  we have introduced the notation
 \begin{equation}
\left< \hat O\right>_{t} \equiv \rm{Tr}\left[\hat O \rho (t)\right].
 \end{equation}
 This can be evaluated as follows
 \begin{eqnarray}
 \Delta E_{\rm tot}^\nu \left[ \rho(t_n^-)\right]&= -\left<\hat  H^\nu -\hat  H_{\rm d} - \hat  H_{\rm r} \right>_{t_n^- }  \nonumber \\
 &= -\left<\hat  H^\nu\right>_{t_{n-1}^+}  +\left<\hat  H_{\rm d} + \hat  H_{\rm r } \right>_{t_n^-} \nonumber\\
&= -\left< \hat  H_{\rm d} + \hat  H_{\rm r }  \right>_{t_{n-1}^+}  +\left< \hat  H_{\rm d} +\hat  H_{\rm r } \right>_{t_n^- } \nonumber\\
 &\equiv  \Delta E_{\rm d}^\nu+\Delta E_{\rm r }^\nu.
 \label{eq:energyChange}
 \end{eqnarray}
 From line one to line two we have used that the total system evolves under a conservative time evolution in the interval $(t_{n-1}^+,t_{n}^-)$. Line two is equal to line three as there are no dot-reservoir coherences at time $t_{n-1}^+$. Finally in line four, we have defined the energy differences corresponding to the dot and reservoir subsystem, respectively, thus $\Delta E_{\rm d}^\nu\equiv\left< \hat  H_{\rm d} \right>_{t_n^- }  -\left< \hat  H_{\rm d}   \right>_{t_{n-1}^+} $, and accordingly for $\Delta E_{\rm r }^\nu $.
 
 Equation~\eqref{eq:energyChange} is an interesting result, as it relates the change of energy induced by the measurement at time $t= t_n$ with the energy-conserving time evolution in the preceding time interval $(t_{n-1}^+,t_{n}^-)$. We emphasize that this result is exact and holds even for more complicated Hamiltonians under the assumption that all system-reservoir coherences vanish due to the projective measurement. 
 
 If we consider  a stationary state which is characterized by $\rho_{\rm d,\rm d}(t_{n-1}) =\rho_{\rm d,\rm d}(t_{n}) $, we find for the averaged energy change
 \begin{eqnarray}
\Delta E_{\rm {fb}}  = \sum_{\nu=\rm{E,F} } p_{\rm{s},\nu} \Delta E_{\rm{ tot} }^\nu \left[ \rho_{\rm s}^\nu (t_n^-)\right],
 \end{eqnarray}
 where $ \rho_{\rm s}^\nu (t_n^-)$ is the density matrix if the measurement outcome at time $t_{n-1}$ has been $\nu$. The corresponding probability is denoted by $p_{\rm{s},\nu}$. In the stationary state, the averaged dot energy is constant at times $t_n$, so that we find
  \begin{eqnarray}
\Delta E_{\rm {fb}} = \sum_{\nu=\rm E,\rm F} p_{\rm{s},\nu} \rm{Tr} \left\lbrace \hat H_{\rm r} \left[\rho_{\rm s}^\nu (t_n^-)-\rho_{\rm s}^\nu (t_{n-1}^+)\right] \right\rbrace .
 \end{eqnarray} 
This is the energy entering the reservoir during the interval $(t_{n-1}^+,t_n^-)$ averaged over the measurement results $\nu$ in the stationary state. 
  Using the DCG method, we can thus obtain the feedback energy $\Delta E_{\rm{fb}}$ by deriving the MGF 
   \begin{eqnarray}
  \Delta E_{\rm{fb}} &= -\rm i \sum_{\alpha=\rm {L,R} } \frac{d}{d \zeta_\alpha} \left. M(\tau,\underline \xi) \right|_{\underline \xi=0} \nonumber \\
  &\equiv \Delta E_{\rm L} + \Delta E_{\rm R} ,
  \end{eqnarray}
  where we have finally partitioned the total energy change into energies entering the left $ \Delta E_{\rm L} $ and right $  \Delta E_{\rm R}$ reservoirs  within the feedback period $\tau$. We note, that this result is consistent with the first law of thermodynamics.
  
  In Fig.~\ref{fig:powerDiagram}(b) we depict the gain $G$ in the same regime as in (a) where we generate power, $\overline P>0$. For regions where  $\Delta E_{\rm {fb}}>0$ we use a color code. For a clear representation, we restrict the range to $G\in(0,40)$.   In the blue regions, we find a negative feedback energy $\Delta E_{\rm {fb}}<0$, thus, the total system energy decreases due to the measurement and the demon  does not perform work on the system but extracts work. It is thus most profitable to operate the system in this region. 
  
   Close to the transition at $\Delta E_{\rm {fb}}=0$ the gain diverges. This line represents the original idea of the Maxwell demon that due to a energy conserving action of the demon (measurement, opening and closing the door)  one can generate a thermal gradient or increase a chemical potential bias. However, we emphasize that even though  in the quantum regime the measurement does not change the energy balance, it changes the state of the system. This is in contrast to the classical Maxwell demon, where the measurement leaves the system state unaffected.
   
   In principle, one could argue that a negative feedback energy  $\Delta E_{\rm {fb}}<0$ could be stored or used by a  smart demon for another application. However, a detailed discussion of this issue  could become possible when specifying the measurement apparatus~\cite{Schaller,Deffner2013,Bozkurt2017}.
  
  \subsection{Heat, entropy and information efficiency }
  
  Next we discuss the heat flow and the thermodynamic consistency. In the stationary state, the change of heat entering the reservoirs within a feedback period $\tau$ reads
  \begin{eqnarray}
 \Delta Q &= \Delta Q_{\rm L} +\Delta Q_{\rm R},   \label{eq:defHeat} \\
 \Delta  Q_\alpha &= \Delta E_\alpha - \mu_\alpha \Delta n_\alpha \nonumber .
  %
  \end{eqnarray}
  The heat is depicted in Fig.~\ref{fig:powerDiagram}(c).  The solid lines represent sets where the total heat change in the reservoirs vanishes $\Delta Q=0$. These lines resemble roughly the zero power line $\overline P=0$. A power generation is thus correlated by an overall loss of heat in the reservoirs.

  \begin{figure}[t]
    \centering
    \includegraphics[width=.5\linewidth]{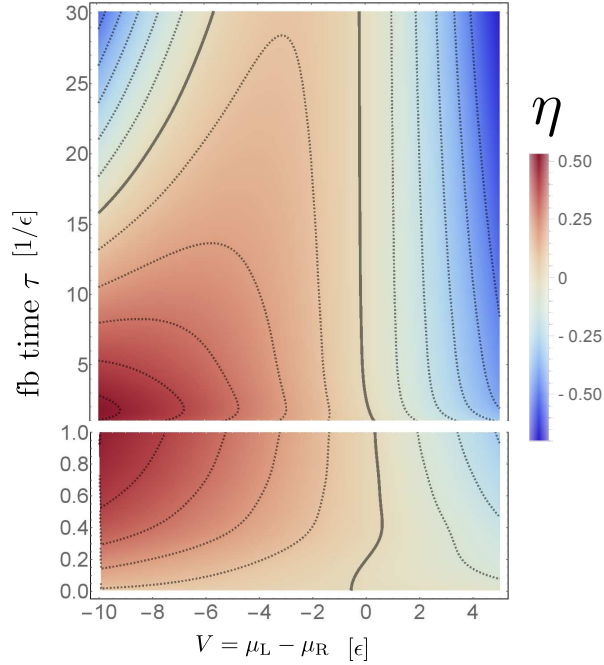}
    \caption{ (a) Information efficiency $\eta$ as defined in Eq.~\eqref{eq:efficacy}. The parameters are as in Fig.~\ref{fig:powerDiagram}.
    \label{fig:efficacyDiagram}
    } 
    \end{figure}

The second law of thermodynamics says that on average the total entropy increases in time in the absence of feedback processes. In a stationary state, this relation reads
 \begin{equation}
 \Delta S_{\rm i}= \Delta S+\Delta S_{\rm e}\geq 0,
 \label{eq:secondLaw}
 \end{equation}
 where $\Delta S_{\rm i}$ denotes the entropy production and $\Delta S_{\rm e}$ is the entropy change in the reservoirs within a time interval $\tau$. The change of entropy in the system (i.e., quantum dot) is given by
 \begin{eqnarray}
 \Delta S &=S(t+\tau)-S(t) \\
 S(t)&=- k_{\rm B} \rm{Tr} \left[\rho_{\rm d}(t) \ln \rho_{\rm d}(t)\right] . \nonumber
 \end{eqnarray}
  In Ref.~\cite{Esposito2010}, Esposito and coworkers have  derived a general relation for the entropy change which is valid for  arbitrary  system-reservoir setups.
 Under the assumption of a product initial state as in Eq.~\eqref{eq:initialState} and a unitary time evolution, the entropy change in the reservoirs reads 
 \begin{equation}
\Delta S_{\rm e} = \sum_{\alpha} \beta_\alpha \Delta Q_\alpha .
\label{eq:enthropieFlow}
 \end{equation}
 It is  straightforward to generalize this to the feedback protocol considered here. To this end, we consider the change of entropy conditioned on the measurement outcome at time $t=t_{n}$ within the subsequent feedback period $t\in \left( t_{n}^+,t_{n+1}^- \right)$. For both measurement outcomes the second law in Eq.~\eqref{eq:secondLaw} together with Eq.~\eqref{eq:enthropieFlow} is fulfilled separately, so that we find for the measurement-averaged entropy change
 \begin{equation}
 \Delta \tilde S + \Delta \tilde S_{\rm e }\equiv
 \sum_{\nu=\rm E,F} p_\nu \Delta S^{(\nu)} + \sum_{\nu=\rm E,F} p_\nu \Delta S_{\rm e }^{(\nu)} \geq 0.
 \label{eq:averagedEnthropy}
 \end{equation}
 The action of the measurement  is to delete the entropy of the system by exactly the amount $\mathcal I = -\Delta \tilde S $ during the virtual time interval $(t_{n+1}^-,t_{n+1})$~\cite{Chida2017}. For this reason, one can interpret Eq.~\eqref{eq:averagedEnthropy} in the following way: the amount of entropy reduced by the measurement is not completely transfered to the reservoirs~\cite{Horowitz2014}.  Moreover, it is not hard to prove that $\Delta \tilde S \leq k_{\rm B} \ln 2 $, so that we recover the Landauer principle~\cite{Landauer1961}. 
 
 Equation~\eqref{eq:averagedEnthropy} allows  to define a coefficient which measures how efficient the information is used to decrease the entropy in the reservoir
 \begin{equation}
 \eta = \frac{\Delta \tilde S_{\rm e }}  {\mathcal I }  \leq 1,
 \label{eq:efficacy}
 \end{equation}
 which we denote as information efficiency in the following~\cite{Esposito2012}.
 While $\mathcal{I} <0$, the entropy change in the reservoir $\Delta \tilde S_{\rm e }$ can be both, positive or negative, so that the information efficiency is not bounded from below.  For similar inequalities in other feedback systems we refer to \cite{Sagawa2008,Sagawa2012}.
 
 We depict $\eta$ in Fig.~\ref{fig:efficacyDiagram}. We observe that the information efficiency is bounded by $ \eta\leq 1$, which is a sanity check for the applied DCG method. The information efficiency is rather similar to the heat $\Delta Q$ entering the reservoirs. This is a consequence of the dot occupation, which is approximately  $n_{\rm s}=p_{\rm F} \approx 0.5$ for the considered parameters.

\subsection{Feedback energy and gain in the Zeno regime}
 
 As the general expression for the matter and energy current is rather involved, it is hard to understand under which circumstances the feedback energy is small or even negative. For this reason, we focus on the Zeno regime in the following, where the expressions are simpler. In this regime, the feedback energy reads
 \begin{equation}
 \Delta E_{\rm fb}=  \tau^2 \sum_{\alpha=\rm R,\rm L} \left[ n_{\rm s,0}  \tilde g_{01}^{\alpha,\rm F}    
+  \left(1-n_{\rm s,0}  \right) \tilde g_{10}^{\alpha,\rm E} \right] ,
 \end{equation}
 where $\tilde g_{xy}^{\alpha,\nu}=\left.\partial_{\zeta_\alpha} g_{xy}^{\alpha,\nu} \right|_{\underline \xi=0}$.
 In order to keep the analysis simple, we focus on an extremal feedback case where $\Gamma_{\rm L}^{\rm F}(\omega)=\Gamma_{\rm R}^{\rm E}(\omega) =0$. In most cases we numerically find that the stationary dot occupation is close to $n_{\rm s,0}\approx 0.5$, so that we find a small or negative current if $\tilde g_{01}^{\rm R,\rm F}<\tilde g_{10}^{\rm L,\rm E}$. This relation implies
 \begin{equation}
 	\tilde \Delta \equiv \int d\omega \Gamma_{\rm R}^{\rm F}(\omega)\omega\left[1-f_{\rm R} (\omega)\right] - \Gamma_{\rm L}^{\rm E}(\omega)\omega f_{\rm L} (\omega)<0.
 \end{equation}
 This condition can be met if the  temperatures in the reservoirs are rather high $\beta_\alpha \epsilon\ll 1$, so that  the Fermi functions are rather close to $f_\alpha (\omega)\approx 0.5$ around a broad range around $\omega=\mu_\alpha$. If additionally $\Gamma_L^{\rm E}(\omega)$ is large for large $\omega$ and $\Gamma_{\rm R}^{\rm F}(\omega)$ is large for small $\omega$, the quantity $\tilde \Delta$ and consequently also the feedback energy $\Delta E_{\rm fb}$ can become  rather small or even negative. 
 
 This is exactly the parameter range which we use in Fig.~\ref{fig:powerDiagram}, although we do not work in an extremal feedback limit. There we have chosen a rather high temperature  $T_\alpha=10 \epsilon$. In the parametrization of the spectral densities in Eq.~\eqref{eq:spectralCouplingDensity}, we use $\epsilon_{\rm L}=5\epsilon$ and $\epsilon_{\rm R}=-1\epsilon$.
 
 Furthermore, we can infer  from Eq.~\eqref{eq:timeAveragedBias} that a large bias $V$ results in a large time-averaged power $\overline P$. Although this has a detrimental effect on the averaged matter current $\overline I_{\rm  m}$, the overall effect is indeed a large power as we can see in Fig.~\ref{fig:powerDiagram}(a).

\section{Discussion and conclusions}  
\label{sec:discussionsConclusions}

We have harnessed a DCG approach in order to conveniently describe the dynamics of the SET under the action of the Maxwell demon. We have implemented the demon by a piecewise-constant feedback scheme, where the occupation of the quantum dot is projectively measured with frequency $ 1/ \tau $. The accuracy of the DCG has been tested by benchmarking it with the exact solution in the absence of feedback.
For vanishing feedback times $\tau$, which corresponds to a continuous observation of the system by the demon, we resembled the quantum-Zeno effect by which the current between the reservoirs is blocked. 
Moreover, we found that the power and efficiency are optimized for an intermediate feedback time $\tau$ outside of the quantum-Zeno regime. The performance of the system is also better for a large bias and higher temperatures. With the DCG method we could thus show that there is an intermediate regime between a genuine quantum effect and a classical rate equation dynamics to optimize the performance of a quantum device under dissipative conditions. This seems thus reminiscent to the interplay of quantum and dissipative effects in other transport scenarios~\cite{Xu2016,Xu2016a,Liu2017,Wu2013}.

Furthermore, we have discovered a novel aspect appearing  in the quantum treatment of  Maxwell's demon. Due to the projective measurement of the system, there is a parameter regime where the total system energy \textit{decreases}.  It is the regime where it is most profitable to run the setup. However, whether or not this work can be stored or harnessed to run a third task lies outside the scope of our methods. To approach this question a microscopic implementation of the measurement apparatus would be necessary in contrast to the bare effective description of the projective measurement applied here. A possible and experimentally realistic way would be to describe the measurement process by an adjacent quantum point contact~\cite{Flindt2009,Wagner2017,Fujisawa2006}  or an autonomous feedback setup as investigated in Ref.~\cite{Koski2015,Strasberg2013,Horowitz2014}  .

\section{Acknowledgements}

Financial support by the DFG (SFB~910, GRK~1558, SCHA~1646/3-1, BR~1928/9-1) and the WE-Heraeus foundation (WEH 640) is gratefully acknowledged.
We thank Philipp Strasberg and Javier Cerrillo for constructive discussions.

\section*{References}

\bibliographystyle{unsrt}
\bibliography{transport}

\appendix

\section{Heat entering the reservoirs}

\begin{figure}[h]
    \centering
    \includegraphics[width=0.8\linewidth]{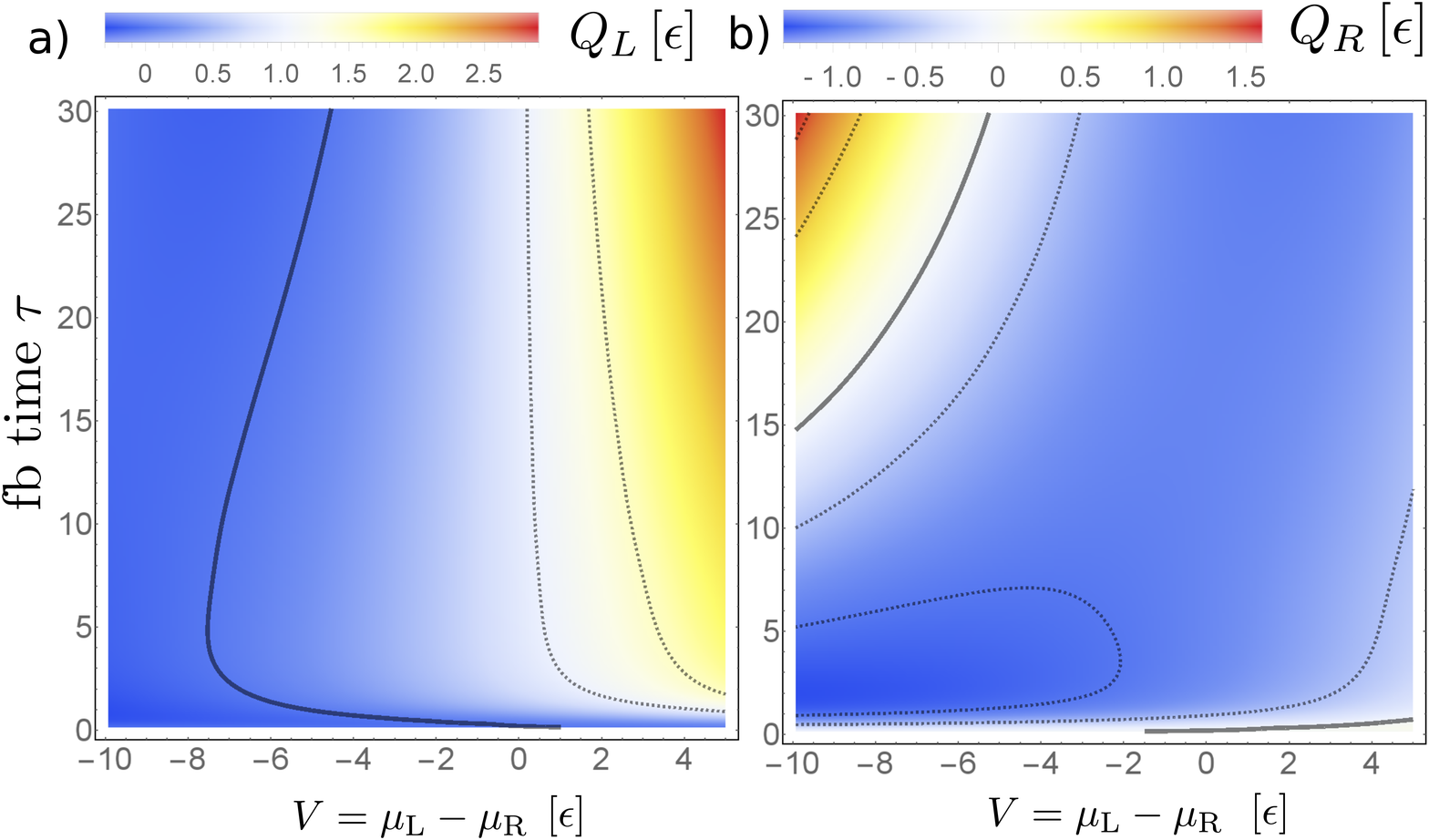}
    \caption{(a) Amount of  heat entering  the left reservoir within one period. (b) Amount of  heat entering  the right reservoir within one feedback period. The parameters are as in Fig.~\ref{fig:powerDiagram}.}
    \label{fig:heatFlows}   
    \end{figure}

In Fig.~\ref{fig:heatFlows}, we depict the amounts of heat transported  to the single  reservoirs $\alpha=\rm R,L$. Overall, the  respective heat amounts  differ strongly from the total heat $Q$ depicted in Fig.~\ref{fig:powerDiagram}(c). In the region, where the transported amount of heat is overall negative $Q<0$, we find that the heat amounts into the reservoirs $Q_\alpha$ is rather small or even negative. If the overall amount of heat is positive $Q>0$, then only either reservoir experiences a strong increase of heat. For $V<0$, the heat in the right reservoir strongly increases and for $V<0$, the heat in the left reservoir strongly increases. 

  \end{document}